\documentclass[iop,apj,useAMS,usenatbib]{emulateapj-rtx4}
\usepackage{amsmath,epsfig,url,natbib}
\usepackage{graphicx,epstopdf,float,color}
\usepackage{courier}
\usepackage[hang,flushmargin]{footmisc}
\usepackage{hyperref}
\hypersetup{colorlinks=true, citecolor=blue, filecolor=magenta,
  urlcolor=cyan, linkcolor=blue}
\def\lsim{\mathrel{\rlap{\lower4pt\hbox{\hskip1pt$\sim$}}
    \raise1pt\hbox{$<$}}}                
\def\gsim{\mathrel{\rlap{\lower4pt\hbox{\hskip1pt$\sim$}}
    \raise1pt\hbox{$>$}}}                

\submitted{}
\begin{document}
\title{{\it Spitzer} Observations of the North Ecliptic Pole}
\author{H. Nayyeri\altaffilmark{1}}
\author{N. Ghotbi\altaffilmark{1}}
\author{A. Cooray\altaffilmark{1}}
\author{J. Bock\altaffilmark{2,3}}
\author{D. L. Clements\altaffilmark{4}}
\author{M. Im\altaffilmark{5}}
\author{M. G. Kim\altaffilmark{5,6}}
\author{P. Korngut\altaffilmark{2,3}}
\author{A. Lanz\altaffilmark{3}}
\author{H. M. Lee\altaffilmark{5}}
\author{D. H. Lee\altaffilmark{6,7}}
\author{M. Malkan\altaffilmark{8}}
\author{H. Matsuhara\altaffilmark{9}}
\author{T. Matsumoto\altaffilmark{10}}
\author{S. Matsuura\altaffilmark{9,10}}
\author{U. W. Nam\altaffilmark{6}}
\author{C. Pearson\altaffilmark{11,12,13}}
\author{S. Serjeant\altaffilmark{12}}
\author{J. Smidt \altaffilmark{14}}
\author{K. Tsumura\altaffilmark{15}}
\author{T. Wada\altaffilmark{9}}
\author{M. Zemcov\altaffilmark{16,2}}

\altaffiltext{1}{Department of Physics and Astronomy, University of
  California Irvine, Irvine, CA 92697, USA}
\altaffiltext{2}{Jet Propulsion Laboratory (JPL), National Aeronautics
  and Space Administration (NASA), Pasadena, CA 91109, USA}
\altaffiltext{3}{Department of Physics, Mathematics and Astronomy,
  California Institute of Technology, Pasadena, CA 91125, USA}
\altaffiltext{4}{Astrophysics Group, Imperial College, Blackett
  Laboratory, Prince Consort Road, London SW7 2AZ, UK}
\altaffiltext{5}{Department of Physics and Astronomy, Seoul National
  University, Seoul 151-742, Korea}
\altaffiltext{6}{Korea Astronomy and Space science Institute, Daejeon
  34055, Korea}
\altaffiltext{7}{University of Science and Technology, Daejeon 34143, Korea}
\altaffiltext{8}{Department of Physics and Astronomy, University of California, Los Angeles, CA 90095-1547, USA}
\altaffiltext{9}{Department of Space Astronomy and Astrophysics, the
  Institute of Space and Astronautical Science, Japan Aerospace
  Exploration Agency, Sagamihara, Kanagawa 252-5210, Japan}
\altaffiltext{10}{School of Science and Technology, Kwansei Gakuin
  University, Sanda, Hyogo 669-1337, Japan}
\altaffiltext{11}{RAL Space, Rutherford Appleton Laboratory, Chilton, Didcot, Oxfordshire
OX11 0QX, UK}
\altaffiltext{12}{School of Physical Sciences, The Open University, Milton Keynes, MK7
6AA, UK}
\altaffiltext{13}{Oxford Astrophysics, Denys Wilkinson Building, University of Oxford,
Keble Rd, Oxford OX1 3RH, UK}
\altaffiltext{14}{Theoretical Division, Los Alamos National Laboratory,
  Los Alamos, NM 87545, USA}
\altaffiltext{15}{Frontier Research Institute for Interdisciplinary
  Science, Tohoku University, Sendai, 980-8578, Japan}
\altaffiltext{16}{Center for Detectors, School of Physics and
  Astronomy, Rochester Institute of Technology, Rochester NY 14623, USA}

\journalinfo{Accepted to the Astrophysical Journal Supplement Series}
\begin{abstract}
We present a photometric catalog for {\it Spitzer}
Space Telescope warm mission observations of the North Ecliptic Pole (NEP; centered
at $\rm R.A.=18^h00^m00^s$, $\rm Decl.=66^d33^m38^s.552$). The
observations are conducted with IRAC in 3.6\,$\mu$m and 4.5\,$\mu$m
bands over an area of 7.04\,deg$^2$ reaching 1$\sigma$ depths of
1.29\,$\mu$Jy and 0.79\,$\mu$Jy in the 3.6\,$\mu$m and 4.5\,$\mu$m bands
respectively. The photometric catalog contains 380,858 sources with
3.6\,$\mu$m and 4.5\,$\mu$m band
photometry over the full-depth NEP mosaic. Point source
completeness simulations show that the catalog is 80\% complete down to
19.7\,AB. The accompanying catalog can be utilized in constraining the
physical properties of extra-galactic objects, studying the AGN
population, measuring the infrared colors of stellar objects, and
studying the extra-galactic infrared background light.

\end{abstract}


\section{Introduction}

A statistical understanding of galaxy properties could be achieved by
measuring the number counts of observed sources as a function of brightness in
wide-area surveys \citep{Jones1991, Pozzetti1998, Yasuda2001,
  Hatsukade2011, Valiante2016, Geach2017, Hemmati2017}. This has been
utilized successfully in the optical and near-infrared bands to study
galaxy mass assembly and star-formation activity originating from
stellar emission (e.g. \citealp{Gardner1993, Fontana2014}). Infrared observations of
extragalactic sources has been made possible by the first generations of upper
atmosphere probes and space missions \citep{Harwit1966,
  Neugebauer1969, Neugebauer1984,
  Kessler1996}, providing the first studies of non-stellar emission \citep{Saunders1990,
  Genzel1998}. In particular, all-sky observations by the Infrared Astronomical Satellite ({\it
  IRAS}; \citealp{Neugebauer1984}) and Wide-Field Infrared Survey
Explorer ({\it WISE}; \citealp{Wright2010}) provided the first dust
maps \citep{Schlegel1998} and paved the way for detailed studies of
populations of infrared-bright galaxies \citep{Soifer1984, Genzel1998, Calzetti2000,
  Eisenhardt2012, Wu2012, Bridge2013}.

 The {\it Spitzer} Space Telescope \citep{Werner2004} revolutionized
 studies of galaxy evolution by making crucial
 observations in the infrared revealing dust and stellar
 components in astronomical objects \citep{Perez2005, Draine2007,
   Magnelli2011}. {\it Spitzer} deep and wide-field infrared
 observations over the past decade
 have produced a unique dataset \citep{Dickinson2003, Lonsdale2003, Sanders2007,
   Ashby2009, Ashby2013, Ashby2015} that has been used to study galaxy
 formation and evolution across a wide range of redshift and physical
 properties \citep{Lacy2004, Lefloch2005, Brandl2006,
   Papovich2006}.

\begin{figure}
\centering
\includegraphics[trim=0cm 0cm 0cm 0cm,scale=0.2]{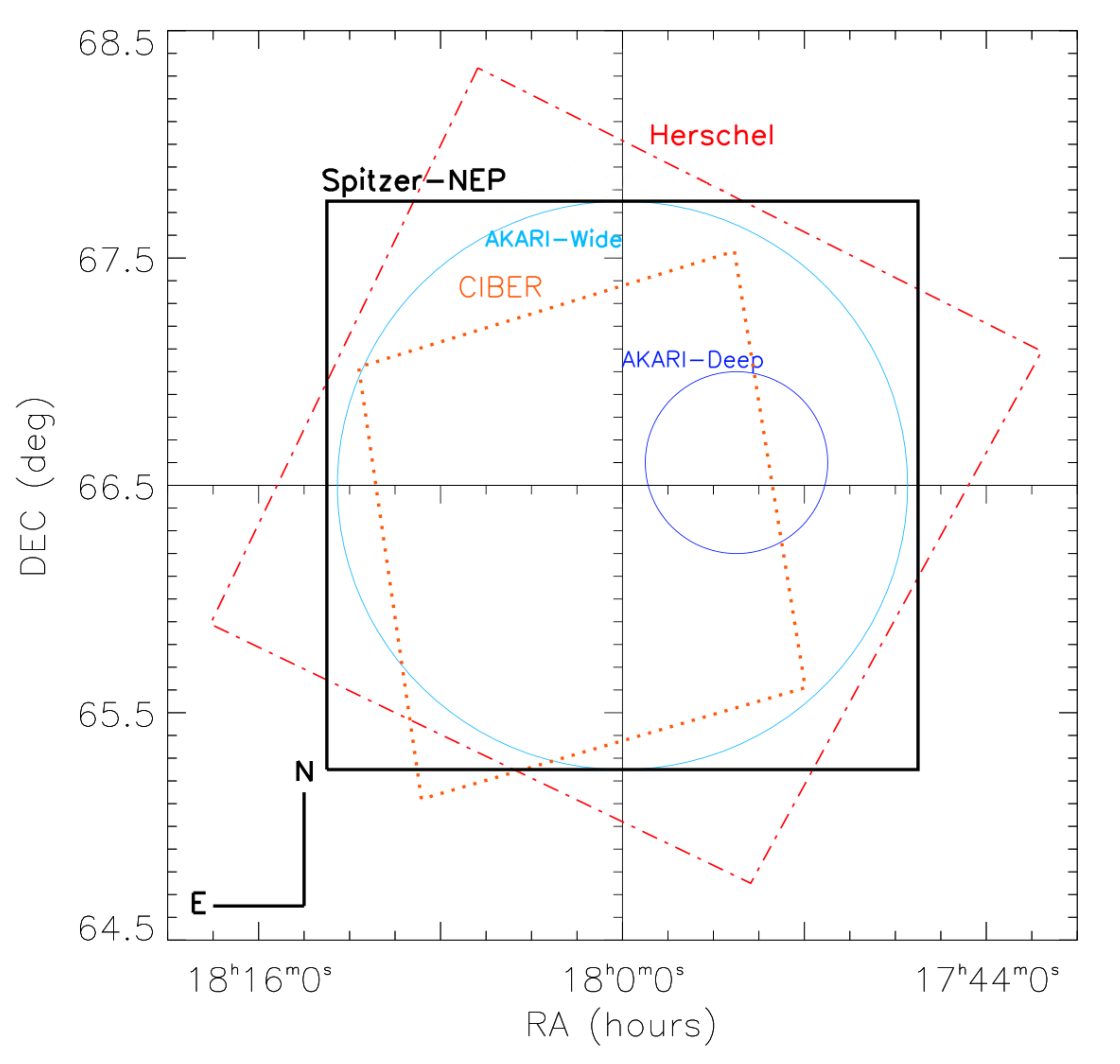}
\caption{The areal coverage of the {\it Spitzer} North Ecliptic Pole (centered
at $\rm R.A.=18^h00^m00^s$, $\rm Decl.=66^d33^m38^s.552$) compared to
that of AKARI \citep{Murakami2007} in the Deep
  \citep{Matsuhara2006,Murata2013} and Wide \citep{Lee2009} fields,
  CIBER \citep{Bock2013, Zemcov2014} and {\it Herschel}
  \citep{Pearson2017}. The {\it Spitzer} observations were chosen to maximize
  the sky coverage by all these surveys providing a complete census of
the infrared SEDs of galaxies.}
\label{fig:Fig1}
\end{figure}

In this work, we present catalogs of stellar and galactic objects
identified from {\it Spitzer} Infrared Array Camera (IRAC;
\citealp{Fazio2004}) observations of the North Ecliptic Pole
(NEP). The NEP (centered at $\rm R.A.=18^h00^m00^s$, $\rm
Decl.=66^d33^m38^s.552$) is the natural extragalactic deep field and
has one of the deepest observations by several space observatories,
including {\em Planck} and AKARI \citep{Serjeant2012}. It further has
extensive observations from the X-ray to the millimeter
\citep{Kollgaard1994, Gioia2003, Lee2007, Jeon2010, Jeon2014,
  Krumpe2015} with additional
observations by {\it WISE} \citep{Jarrett2011}. The field was
specifically chosen to match the
  observations by the Cosmic Infrared Background Experiment (CIBER;
  \citealp{Bock2013, Zemcov2014}) to study the extragalactic background light
  fluctuation \citep{Cooray2004, Matsumoto2005, Cooray2012, Zemcov2014, Matsuura2017}. 
Figure \ref{fig:Fig1} shows the sky coverage of the
 {\it Spitzer} NEP observations compared to the other infrared
 missions of the north ecliptic pole
 \citep{Murakami2007,Matsuhara2006,Murata2013,Lee2009,Pilbratt2010,Bock2013}. The
{\it Spitzer} observations were designed to maximize overlap with
those of AKARI \citep{Murakami2007} in the Deep
  \citep{Matsuhara2006,Murata2013} and Wide \citep{Lee2009} fields and
  observations by {\it Herschel} \citep{Pearson2017}. 

The {\it Spitzer}/IRAC dataset is complimentary to the already existing
optical/near-infrared \citep{Jeon2010, Jeon2014} data and could be
used in combination with those to constrain the stellar mass function of
galaxies in the NEP. The {\it Spitzer} infrared observations in the
NEP could additionally be utilized (in conjunction with the already
existing X-ray observations; \citealp{Krumpe2015}) to identify AGNs
\citep{Stern2005}. The NEP imaging data are particularly important for
studies of diffuse light background fluctuations arising from individually
un-detected sources directly probing early star-formation
\citep{Cooray2004, Cooray2012, Zemcov2014}.

The paper is organized as follows. In Section 2 we present the imaging
data and compilation for the {\it Spitzer} NEP field. Section 3
provides details on our source catalog and photometry estimation.
We discuss our data analysis results in Section 4 and summarize our
findings in Section 5. Throughout this paper we assume a standard cosmology with
$H_0=70\,\text{kms}^{-1}\text{Mpc}^{-1}$, $\Omega_m=0.3$ and
$\Omega_\Lambda=0.7$. All magnitudes are in the AB system where
$\text{m}_{\rm AB}=23.9-2.5\times\text{log}(f_{\nu}/1\mu \text{Jy})$ \citep
{Oke1983}.

\section{Data}

The North Eclipctic Pole, centered at $\rm R.A.=18^h00^m00^s$, $\rm
Decl.=66^d33^m38^s.552$, was observed with {\it Spitzer}/IRAC in Cycle
10 (Program ID: 10147; PI: J. Bock). The observations were carried out over three epochs in both 3.6\,$\mu$m
and 4.5\,$\mu$m bands (Table 1). The first two epochs were separated by $\sim
10$ weeks and epoch three was observed $\sim 8$ weeks after epoch
two. The observations in different epochs enables study of the zodiacal
light contamination, which changes throughout the year, for background
fluctuations \citep{Cooray2012}.

Three epochs were considered sufficient for mapping the {\it Spitzer}
NEP, based on previous observations of SDWFS \citep{Ashby2009}, as
this provided reliable power spectra measurements for
  studying the intra-halo light (IHL) and extragalactic background light (EBL)
  \citep{Cooray2012b, Zemcov2014} combined with existing CIBER and
  AKARI data in the field. The mapping strategy involved observations that
were offset by one-third of the IRAC field of view between successive
passes through each group (where we split the field into sixteen
groups) and were dithered on small scales to
maximize the inter-pixel correlation to facilitate
self-calibrations. Furthermore, to ensure that asteroids are reliably
identified, each group is observed in three passes of 30\,sec
each. The time required to obtain a single 30\,sec pass on a group ensures gaps of
at least 2 hours between observations of each sky position. For
typical asteroid motions of 25\,arcsec\,hour$^{-1}$, asteroids will move
1\,arcmin between maps. Since this is much smaller than a typical map
width, three map observations will reliably track asteroid
motions. We further note here that there are very unlikely to be
  any main belt asteroids (with a 25\,arcsec\,hour$^{-1}$ motion) in the
  Ecliptic Cap, with most being near the Ecliptic Plane. Some
  near-earth asteroids may be present, but these typically have much
  faster motions ($\sim$ several arcmin/hr) and easily identifiable. 

\begin{figure}
\centering
\includegraphics[trim=1cm 0cm 0cm 0cm,scale=0.45]{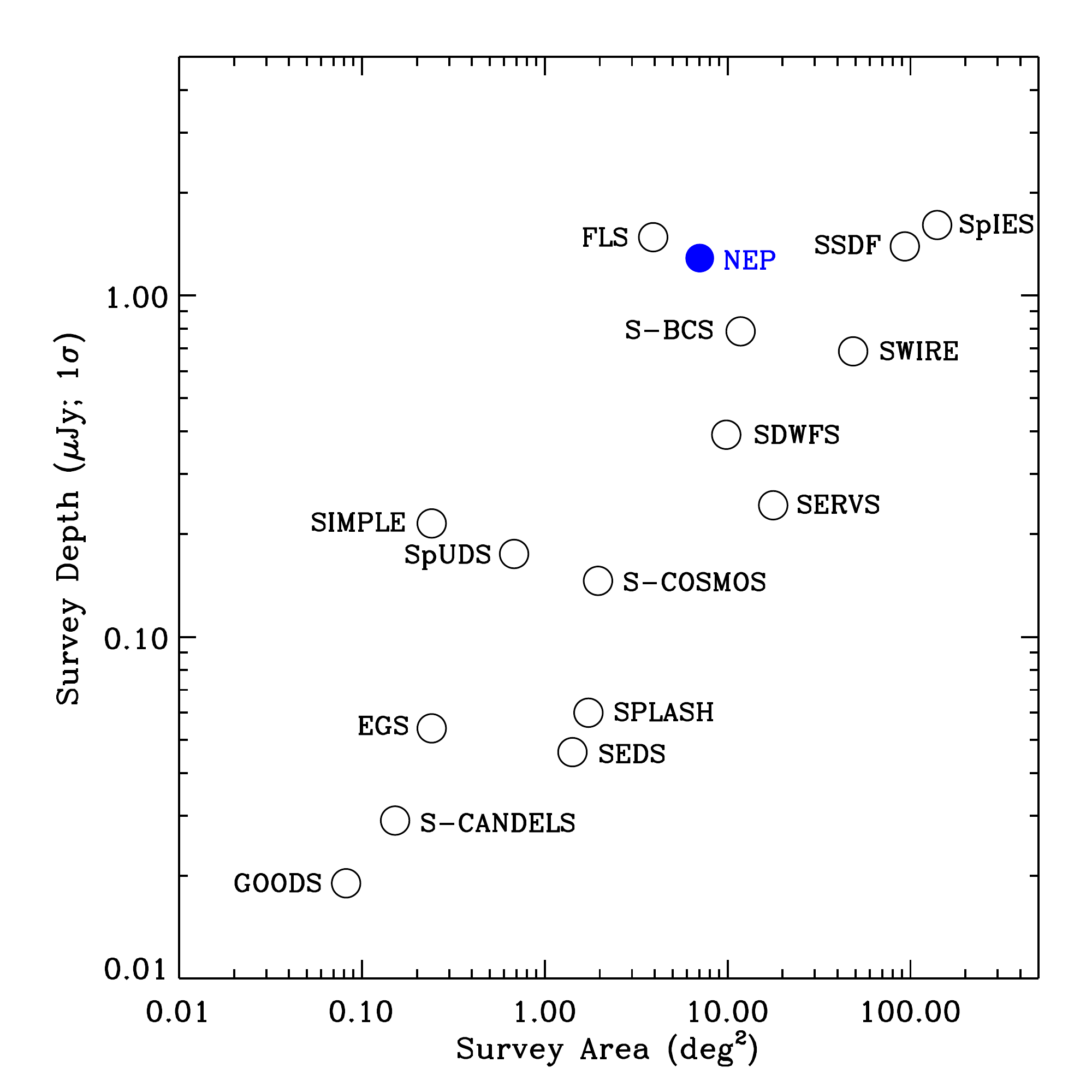}
\caption{Observations depth and sky coverage of the NEP (at the 3.6\,$\mu$m) compared to other {\it
    Spitzer} surveys. These include the Great Observatories Origins
Deep Survey (GOODS; \citealp{Dickinson2003,Giavalisco2004}),
  {\it Spitzer}-Cosmic Assembly Near-infrared Deep Extragalactic Legacy
  Survey (S-CANDELS; \citealp{Ashby2015}), Extended Groth Strip (EGS; \citealp{Davis2007}), {\it Spitzer}
  Extended Deep Survey (SEDS; \citealp{Ashby2013}), {\it Spitzer} Large Area
  Survey with Hyper-Suprime-Cam (SPLASH; \citealp{Capak2013}), S-COSMOS
  \citep{Sanders2007}, {\it Spitzer} Public Legacy Survey of UKIDSS
Ultra-deep Survey (SpUDS; \citealp{Caputi2011}), {\it Spitzer}
IRAC/MUSYC Public Legacy in E-CDFS (SIMPLE; \citealp{Damen2011}), {\it
  Spitzer} Deep Wide-Field Survey (SDWFS; \citealp{Ashby2009}), {\it Spitzer}
  Wide-area Infrared Extragalactic Survey (SWIRE;
  \citealp{Lonsdale2003}), {\it Spitzer} Blanco Cluster Survey
  (S-BCS), First Look Survey (FLS; \citealp{Lacy2005}), {\it Spitzer}
  Extragalactic Representative Volume Survey (SERVS;
  \citealp{Mauduit2012}), {\it Spitzer}-IRAC Equatorial Survey
  (SpIES) and {\it Spitzer} South Pole Telescope Deep Field (SSDF;
  \citealp{Ashby2013b}). The NEP covers an area of 7.04\,deg$^2$ reaching
a 1$\sigma$ depth of 1.29\,$\mu$Jy (measured over an aperture of $4^{\prime\prime}$).}
\label{fig:Fig2}
\end{figure}

Basic Calibrated Data (BCD) and associated
files were directly downloaded from the {\it Spitzer} Science Center
after acquisition. This corresponds to 3936 tiles in each epoch in each
of the bands (except epoch 2 which observed 3854 tiles in 3.6\,$\mu$m;
see Table 1) with 23.6\,sec and 26.8\,sec exposures per tile at 3.6\,$\mu$m and 4.5\,$\mu$m,
respectively. This resulted in a continuous {\it Spitzer} map of the
NEP field covering an area of 7.04\,deg$^2$ corresponding to a total
integration time of $\rm \sim 165\,h$. Figure~\ref{fig:Fig2} shows the
NEP area coverage and depth compared to other {\it Spitzer} surveys.  

\begin{table*}
\begin{center}
\caption{Observations Summary$^\dagger$.}
\begin{tabular}{ccccc}
\hline
\hline
Epoch & Observation Dates & No. BCDs 3.6\,$\mu$m/4.5\,$\mu$m &
                                                                 Exposure
                                                                 Time$^a$
                                                                 3.6\,$\mu$m/4.5\,$\mu$m
  & Exposure time per pixel$^b$ 3.6\,$\mu$m/4.5\,$\mu$m \\ 
 & & & (hours) & (seconds) \\
\hline
1 & May 06-09, 2014 & 3936/3936 & 25.8/29.3 & 106/120 \\
2 & Jul 18-24, 2014 & 3854/3936 & 25.3/29.3 & 104/119 \\
3 & Sep 08-14, 2014 & 3936/3936 & 25.8/29.3 & 107/121 \\
\hline
Total & $-$ & 11808/11726 & 76.9/87.9 & 282/331$^c$ \\
\hline

\end{tabular}
\end{center}
\footnotesize
$^\dagger$: Cycle 10 program {\it Spitzer}/IRAC observations (PID:
10147). $^a$: Total exposure time of each mosaic. $^b$: The average
exposure time per pixel in each mosaic. $^c$: Measured as the
average value from the combined three epoch mosaic exposure maps.
\label{table:Table1} 
\end{table*}

\begin{figure*}[th]
\centering
\includegraphics[trim=3cm 0cm 3cm 0cm,scale=0.25]{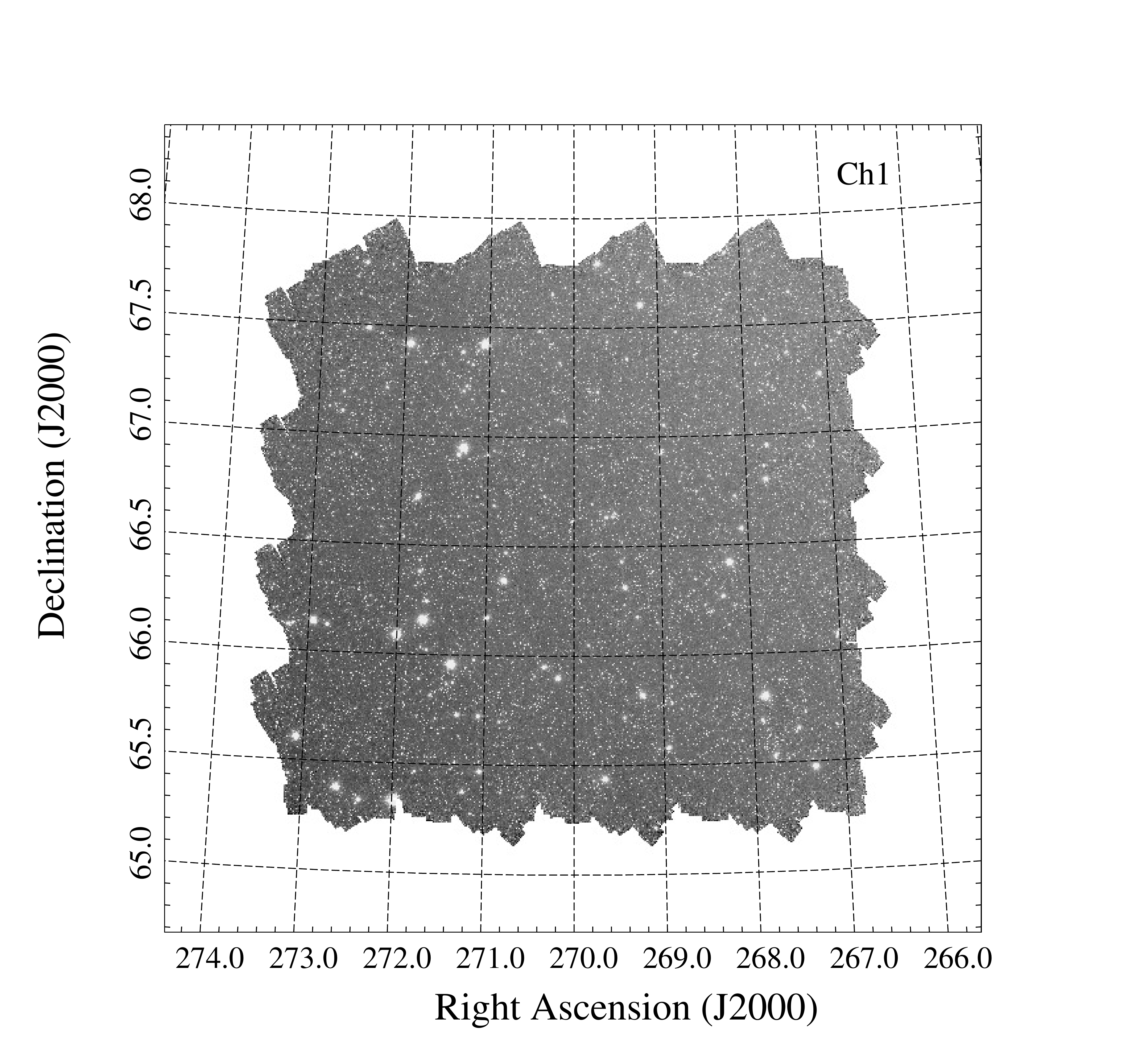}
\includegraphics[trim=3cm 0cm 3cm 0cm,scale=0.25]{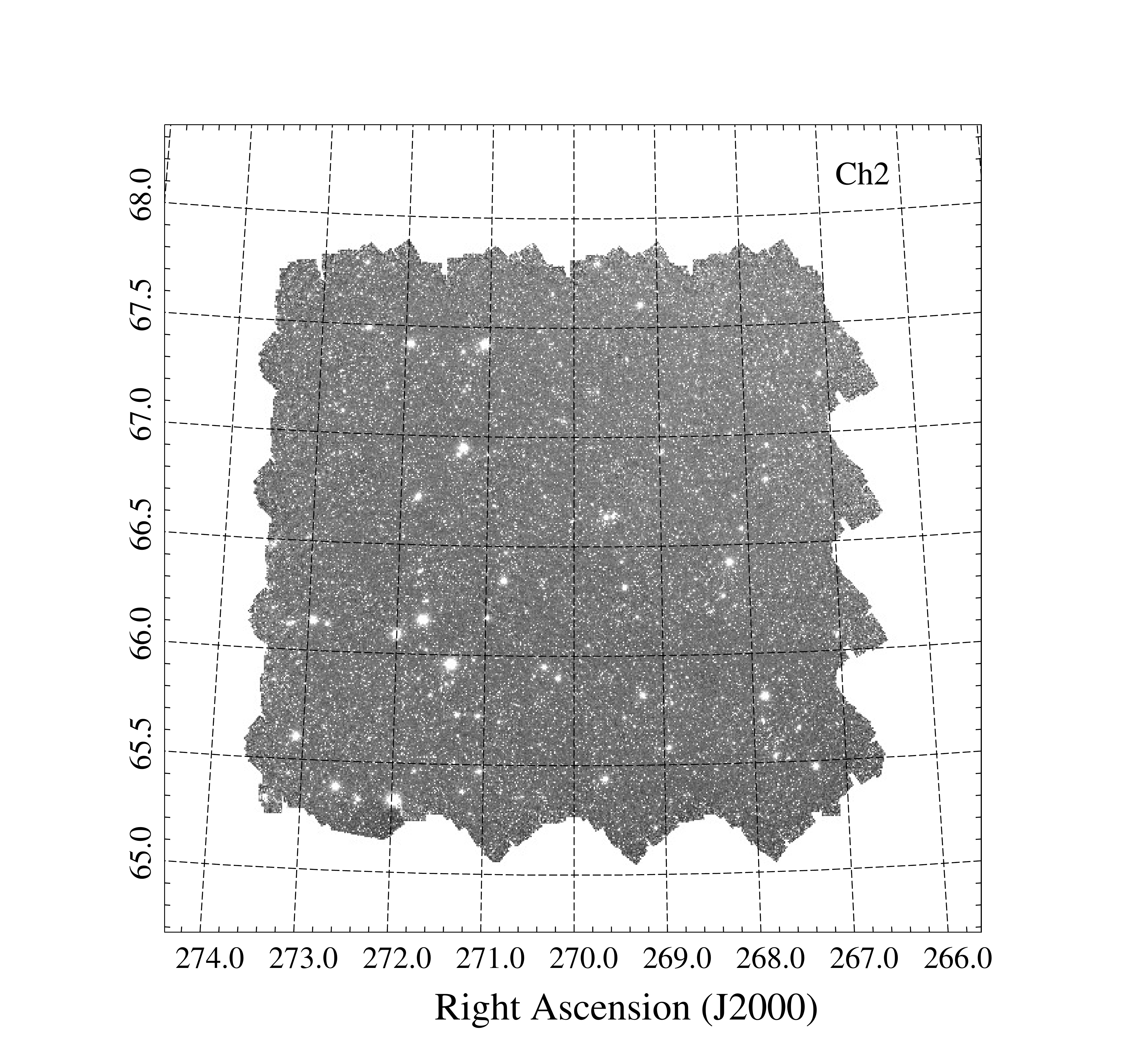}
\caption{The combined three epoch {\it Spitzer} maps of the North
  Ecliptic Pole (NEP) in the 3.6\,$\mu$m (left) and the 4.5\,$\mu$m
  band (right) as part of Cycle 10 observations with total exposure
  time of $\rm \sim 165\,h$ during the warm mission over an area $\rm
  \sim 7.04\,deg^2$.}
\label{fig:Fig3}
\end{figure*}

\begin{figure*}[th]
\centering
\includegraphics[trim=3cm 0cm 3cm 0cm,scale=0.25]{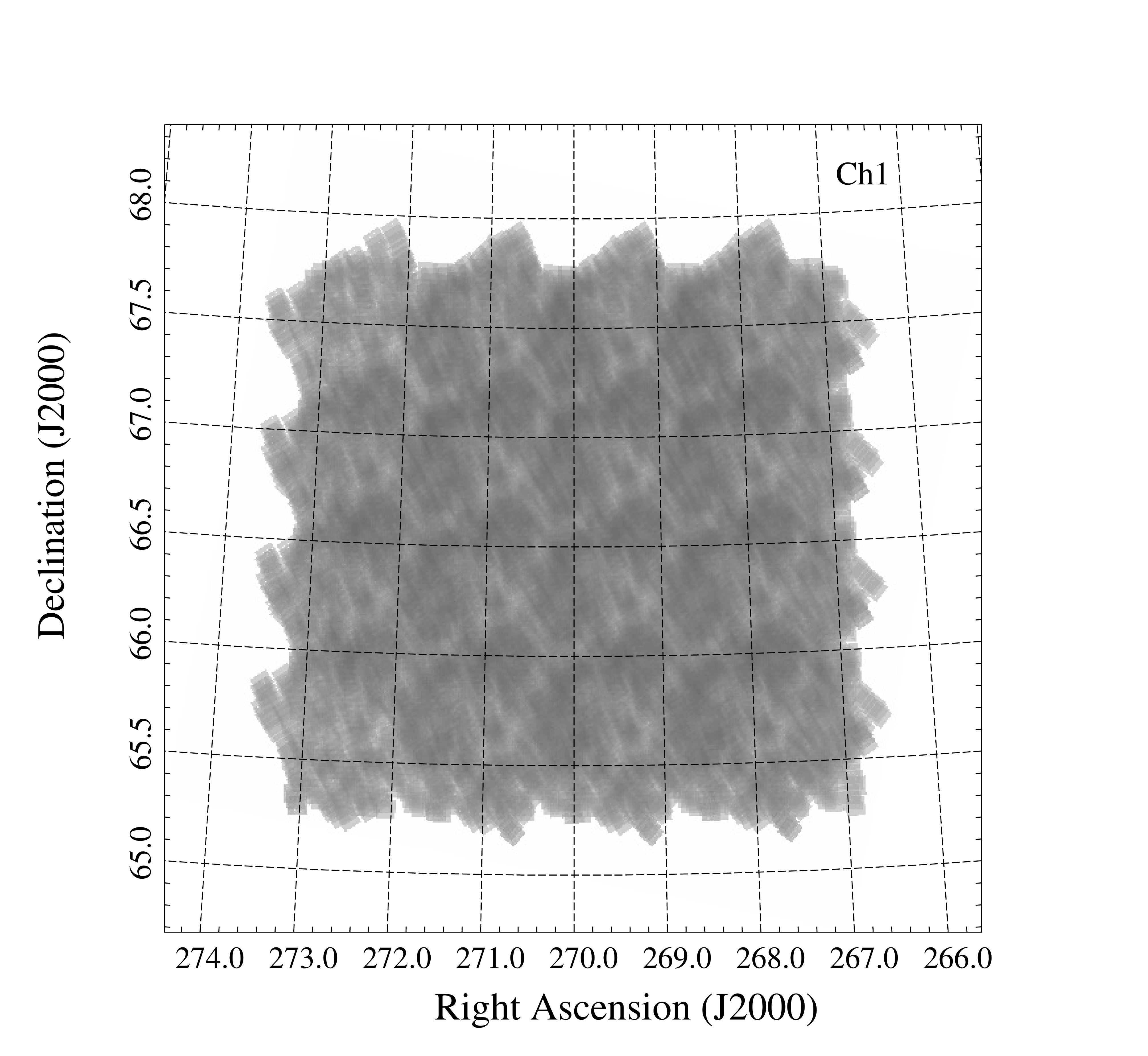}
\includegraphics[trim=3cm 0cm 3cm 0cm,scale=0.25]{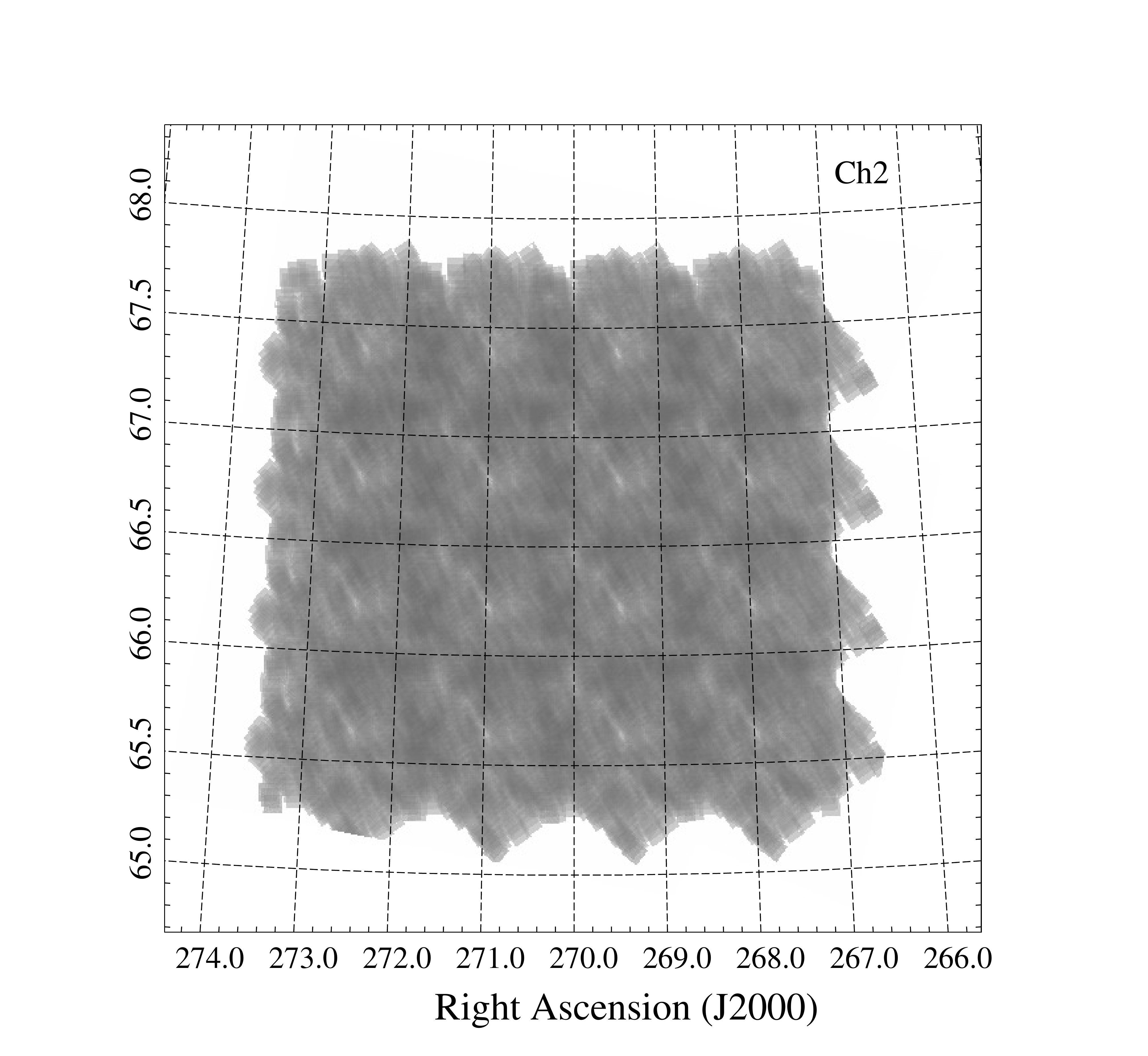}
\caption{The combined three epoch {\it Spitzer} 3.6\,$\mu$m (left) and
  4.5\,$\mu$m (right) coverage maps demonstrating the depth uniformity
  across the field. This consists of total of 11808 and 11726
  individual tiles in 3.6\,$\mu$m and 4.5\,$\mu$m bands respectively over three epochs.}
\label{fig:Fig4}
\end{figure*}

\begin{figure}
\centering
\includegraphics[trim=2cm 0cm 0cm 0cm,scale=0.45]{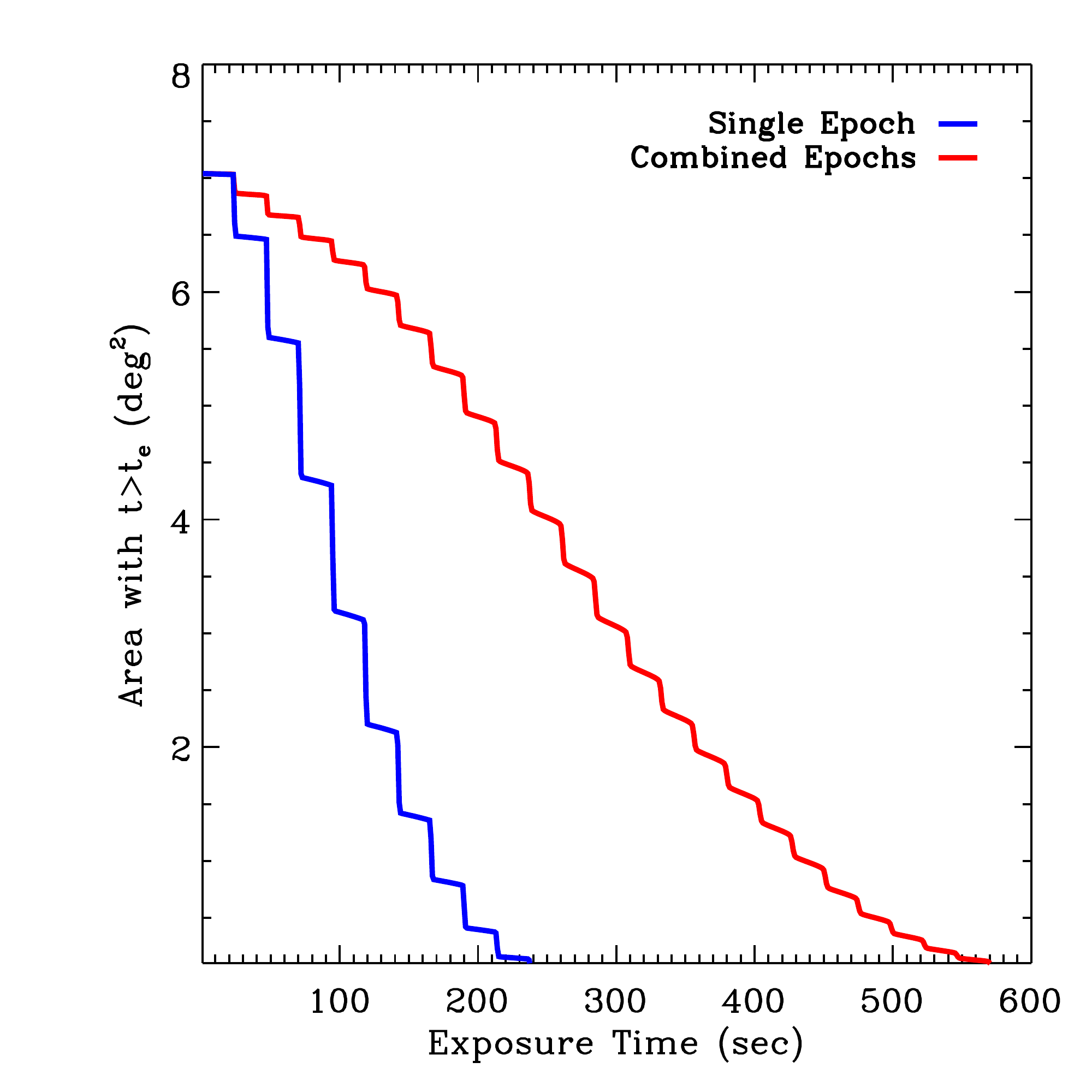}
\caption{Area coverage of the {\it Spitzer} 3.6\,$\mu$m NEP map as a function of
  total exposure time for the combined three epochs (in red) and a
  single epoch (in blue). More than 89\% of the field is covered with
  exposure time of at least 100\,sec in the combined mosaic where that number is 45\% for
  the single epoch observations. Table 1 summarizes the exposure time
  in each epoch.}
\label{fig:Fig5}
\end{figure}

We used {\sc mopex}\footnote{\url{http://irsa.ipac.caltech.edu/data/SPITZER/docs/dataanalysistools/tools/mopex/}}
version 18.5.0 \citep{Makovoz2006} on the corrected-Basic
Calibrated Data (cBCDs) released by the {\it Spitzer} Science Center
to construct the mosaics. We generated combined mosaics for each epoch
and a combined mosaic of all three epochs with {\sc mopex} for the
3.6\,$\mu$m and 4.5\,$\mu$m observations. Figure \ref{fig:Fig3} shows the
combined three epoch mosaics in the 3.6\,$\mu$m and 4.5\,$\mu$m bands. In Summary
{\sc mopex} takes the data frames along with the associated
uncertainty frames and constructs a Fiducial Image Frame (FIF) from the
boundaries of the data frame onto which input observations will be
projected. It then runs a background matching routine on
overlapping frames and performs an image interpolation which maps the
individual frames onto the generated FIF after which a co-added mosaic
image is generated. During the process {\sc mopex} also performs an
outlier rejection routine to identify outlier pixels (such as cosmic
rays) and generates a bad pixel mask which is used in the construction
of the mosaic. Each generated science map is also accompanied
by the corresponding uncertainty and exposure maps calculated by {\sc
  mopex}. We refer the reader to the {\sc mopex}
manual for further details\footnote{\url{http://irsa.ipac.caltech.edu/data/SPITZER/docs/dataanalysistools/tools/mopex/mopexusersguide/}}.
Figure \ref{fig:Fig4} shows the {\sc mopex} generated coverage map of
the NEP for the full depth mosaics in the 3.6\,$\mu$m and 4.5\,$\mu$m bands.
We see that the NEP observations have very uniform depth across the
field in both bands. Figure \ref{fig:Fig5} shows the area coverage of
the 3.6\,$\mu$m map given a minimum exposure time for a single epoch and
the full depth mosaics. We see that more than 89\% of the field (with
area $\rm \sim 6.26\,deg^2$) is covered with an exposure time of at
least 100\,sec in the full depth mosaic whereas only 45\% (with area
$\rm \sim 3.17\,deg^2$) is covered to similar depths in a single
epoch. The individual 3.6\,$\mu$m epochs have 1$\sigma$ depths
of 2.11\,$\mu$Jy, 2.13\,$\mu$Jy and 2.22\,$\mu$Jy respectively in the
three epochs and a depth of 1.29\,$\mu$Jy over the combined full depth
mosaic. The combined full depth mosaic in the 4.5\,$\mu$m band has a 1$\sigma$
depth of 0.79\,$\mu$Jy (measured over an aperture of
4$^{\prime\prime}$ with aperture corrections applied; see
Section 3.2). Table 1 summarizes the {\it Spitzer} NEP observations.

\begin{figure*}
\centering
\includegraphics[trim=2cm 0cm 0cm 0cm,scale=0.47]{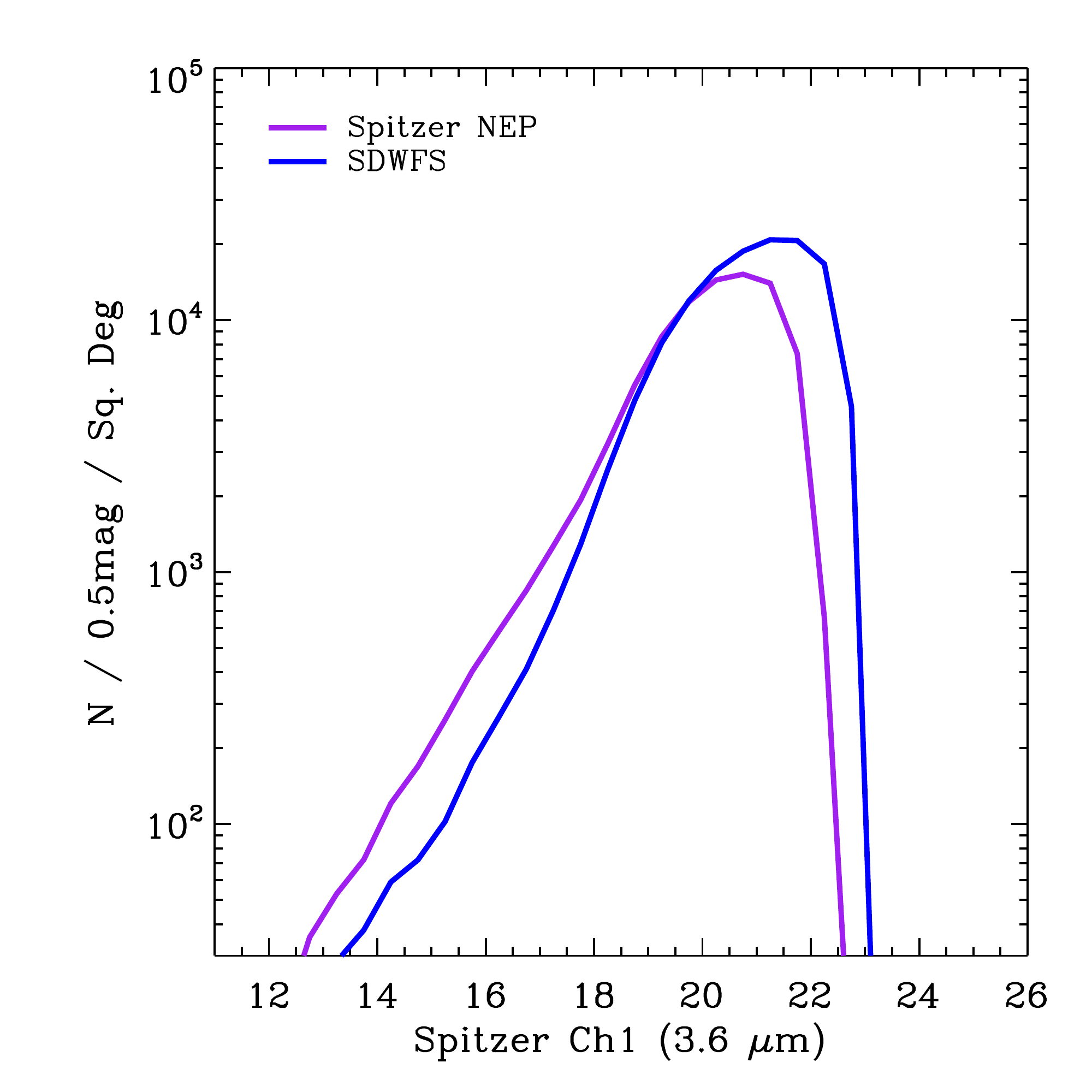}
\includegraphics[trim=2cm 0cm 0cm 0cm,scale=0.47]{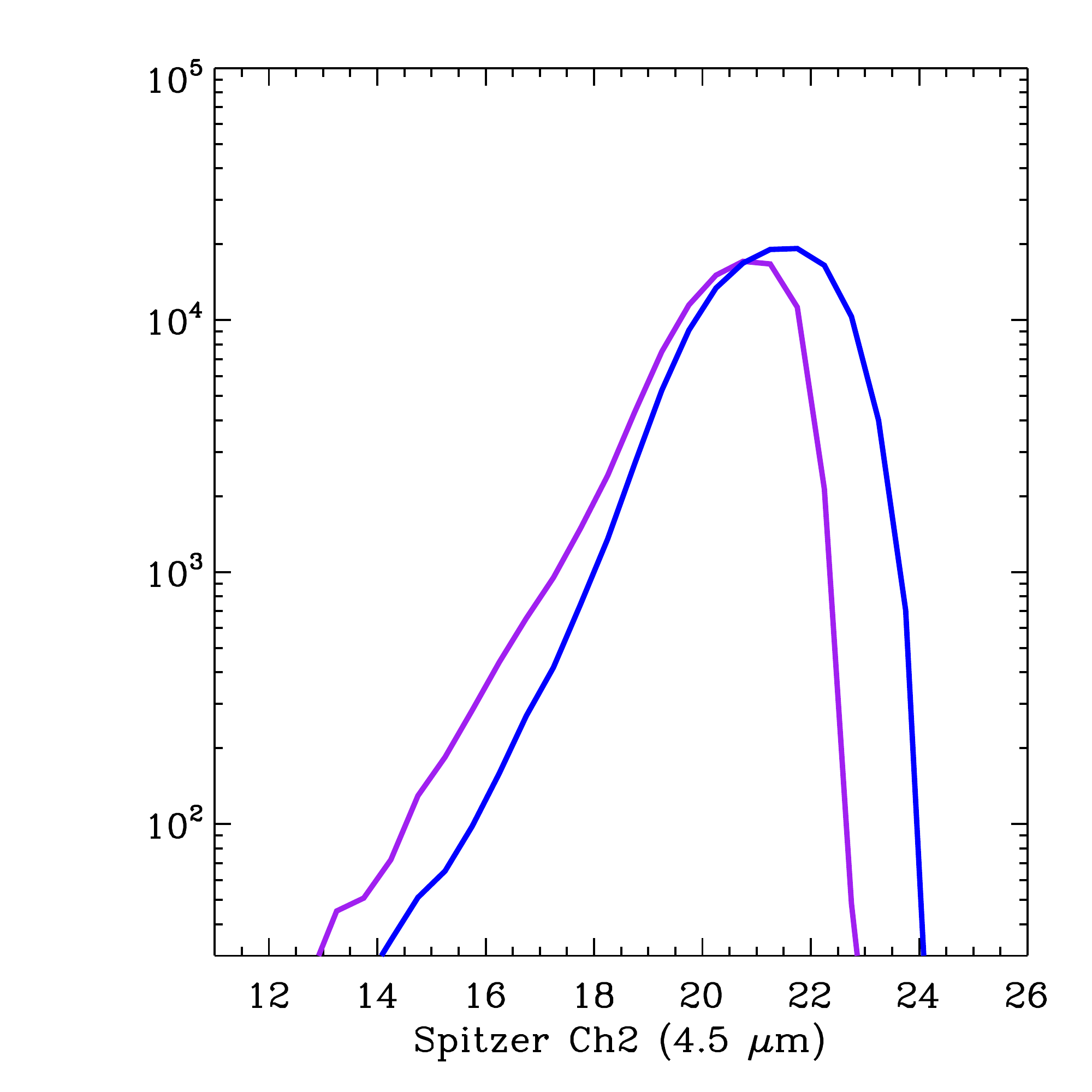}
\caption{Left: Normalized 3.6\,$\mu$m number counts of sources in the {\it
    Spitzer} NEP catalog (magenta) compared to the counts in the SDWFS
  (blue). The SDWFS is a deeper survey (see Figure \ref{fig:Fig2}) and
  this is reflected in the counts at the faint end. On the other had
  the {\it Spitzer} NEP shows a larger number of sources at the bright
end. Right: Number counts of sources in the {\it Spitzer} NEP
4.5\,$\mu$m band compared to the SDWFS \citep{Ashby2009}.}
\label{fig:Fig6}
\end{figure*}

\begin{figure}
\centering
\includegraphics[trim=2cm 0cm 0cm 1.0cm,scale=0.45]{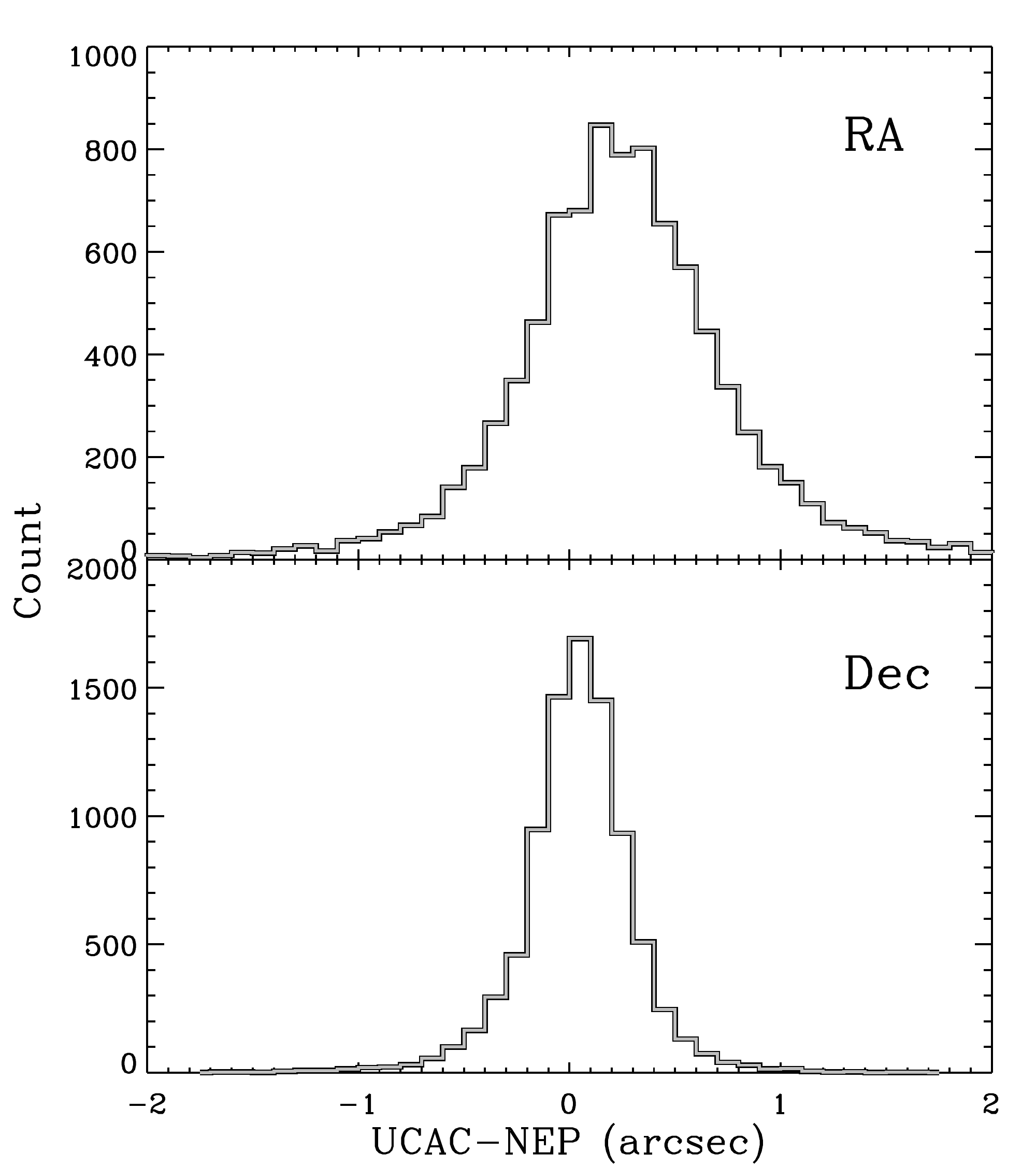}
\caption{Distribution of the difference in right ascension (top) and
  declination (bottom) of the {\it Spitzer} NEP catalog to that of
  USNO CCD Astrograph Catalog (UCAC4; \citealp{Zacharias2013}). The NEP shows an offset of
  0$^{\prime\prime}$.25 in the right ascension (smaller than the {\it
    Spitzer} mosaic pixel) compared to the UCAC
  catalog whereas the Declination distribution is consistent with a
  zero offset. The distributions have standard deviations of
  0$^{\prime\prime}$.51 and 0$^{\prime\prime}$.24 in the right
  ascension and declination respectively that are smaller than that of the IRAC PSF FWHM.}
\label{fig:Fig7}
\end{figure}

\section{Source Catalogs}

\subsection{Source Identification and Photometry}

We used {\sc SExtractor} \citep{Bertin1996} for source identification and
photometry. {\sc SExtractor} is run in dual mode on the combined three
epoch mosaics on the 3.6\,$\mu$m and 4.5\,$\mu$m individually as the
detection bands with photometry extracted in both. The two
catalogs are then merged to form the final NEP catalog. With this
approach, we make sure to include 3.6\,$\mu$m faint sources that are
detected in the 4.5\,$\mu$m and would have been missed by the former
selection alone.  The {\sc mopex} generated mosaics are in units of
$\rm [MJy\,Sr^{-1}]$, which we convert to $\rm [\mu Jy\,pixel^{-1}]$
using the map pixel scale of
$1^{\prime\prime}.2$ before measuring the photometry. We used a minimum
detection area of 3 pixels with a detection threshold of 2$\sigma$ for
source identification. This is chosen to maximize the recovered
sources while minimizing spurious source identification. We identify
380,858 sources over the 7.04\,deg$^{2}$ area of the {\it Spitzer} NEP
map with our detection criteria. We measure
source photometry in AB magnitudes
given the pixel flux units in micro-Jansky and report this in a Kron
radius (the {\sc mag\_auto}), which is photometry over an ellipse with size
and orientation determined from the second moment of light distribution
above the isophotal threshold, and also over two apertures (with
$4^{\prime\prime}$ and $6^{\prime\prime}$ diameter). We further
report the {\sc SExtractor} measured stellarity parameter for each
object which we use later (see Section 4.4) to validate the color of
stars in the field. Figure \ref{fig:Fig6} shows the area-normalized
distribution of the source photometry in the full NEP of the combined
three epochs.

\begin{figure}
\centering
\includegraphics[trim=2cm 0cm 0cm 0cm,scale=0.4]{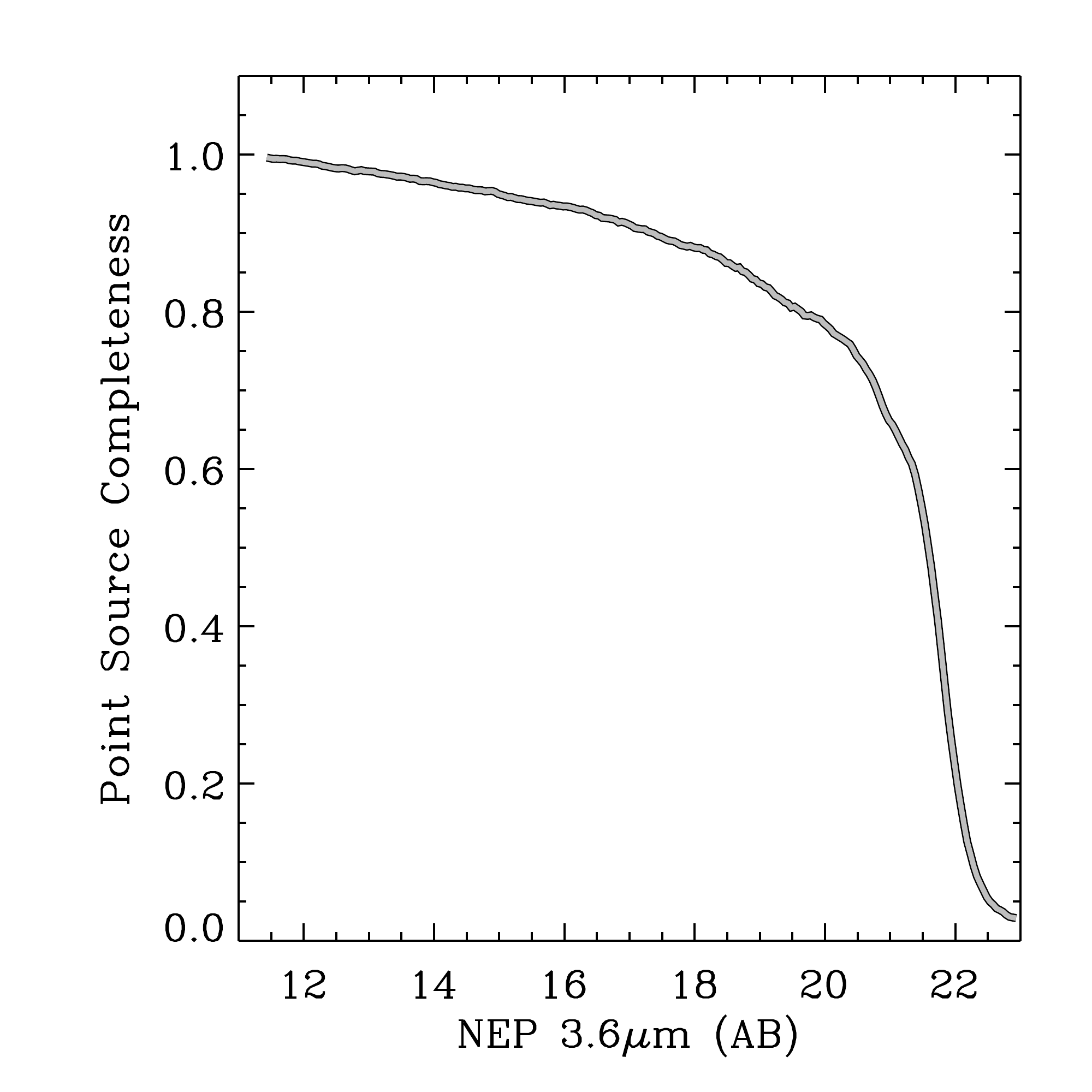}
\caption{Simulated point source completeness fraction at 3.6\,$\mu$m as a function of source
  magnitude. Point sources are generated from the IRAC 3.6\,$\mu$m PSF
and are randomly distributed over the NEP mosaic. The recovered
fraction using the original {\sc SExtractor} parameters, used to
generate the photometric catalog, provides the completeness.}
\label{fig:Fig8}
\end{figure}

\begin{figure*}
\centering
\includegraphics[trim=1cm 0cm 0cm 0cm,scale=0.45]{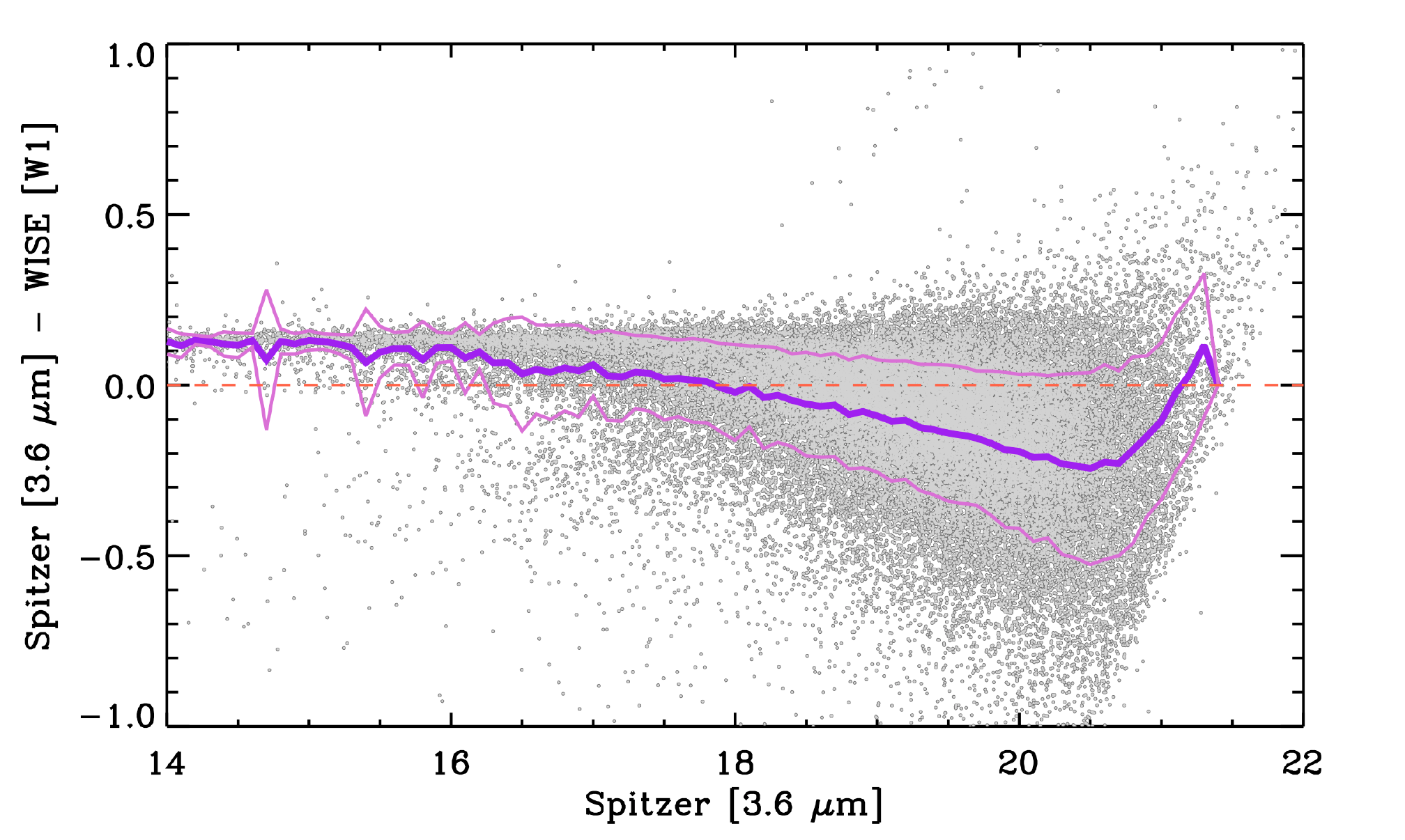}
\includegraphics[trim=1cm 0cm 0cm 0cm,scale=0.45]{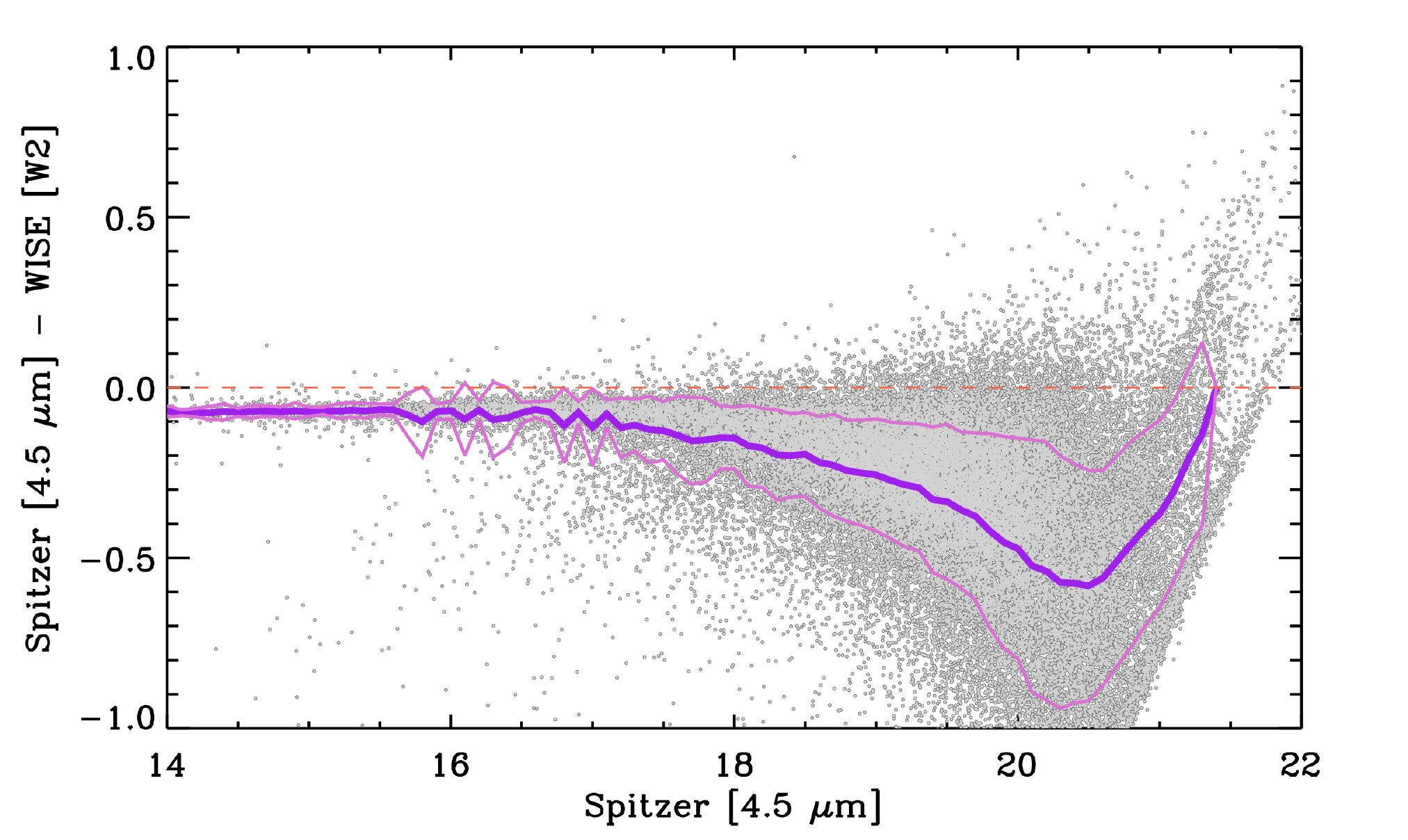}\\
\includegraphics[trim=1cm 0cm 0cm 0cm,scale=0.45]{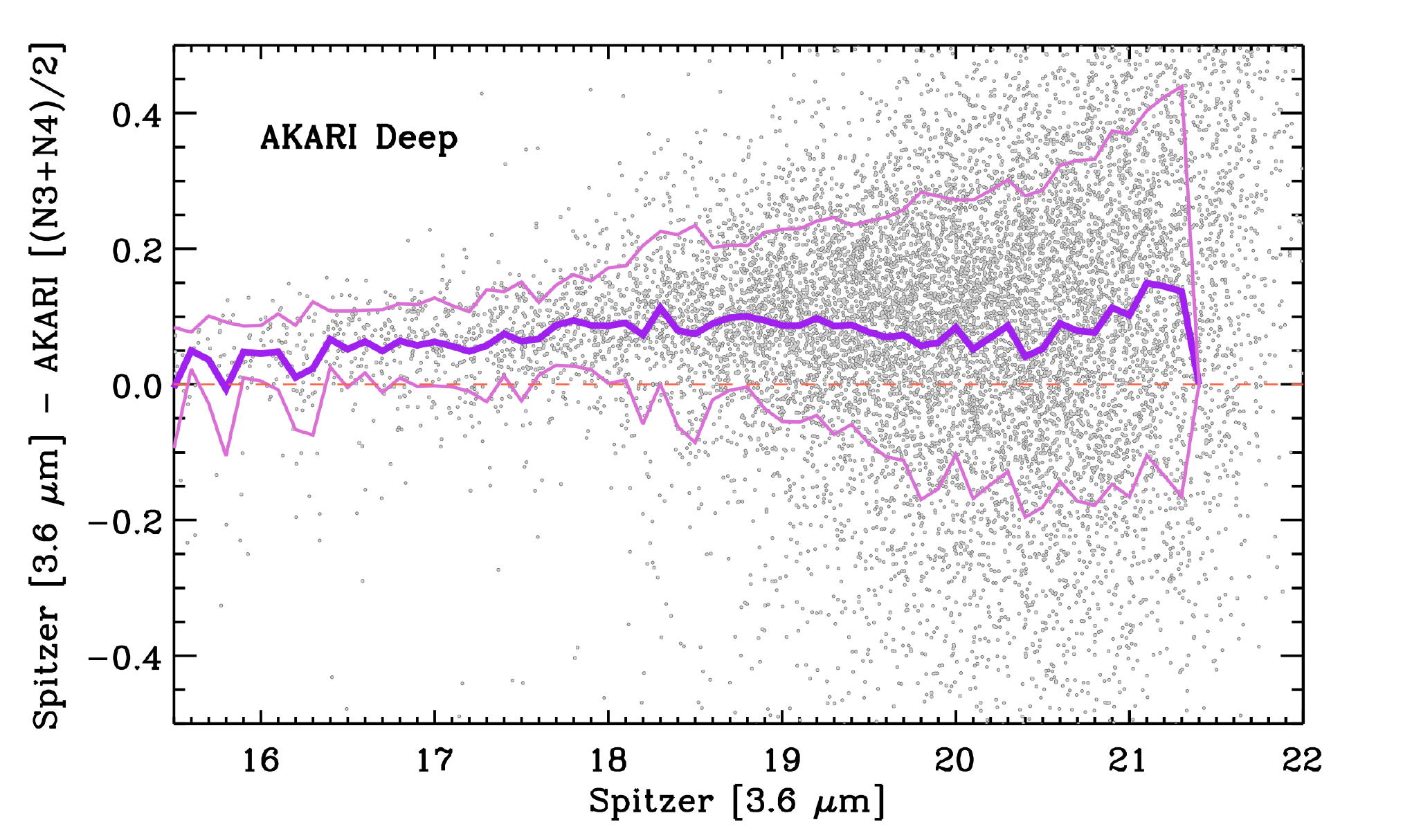}
\includegraphics[trim=1cm 0cm 0cm 0cm,scale=0.45]{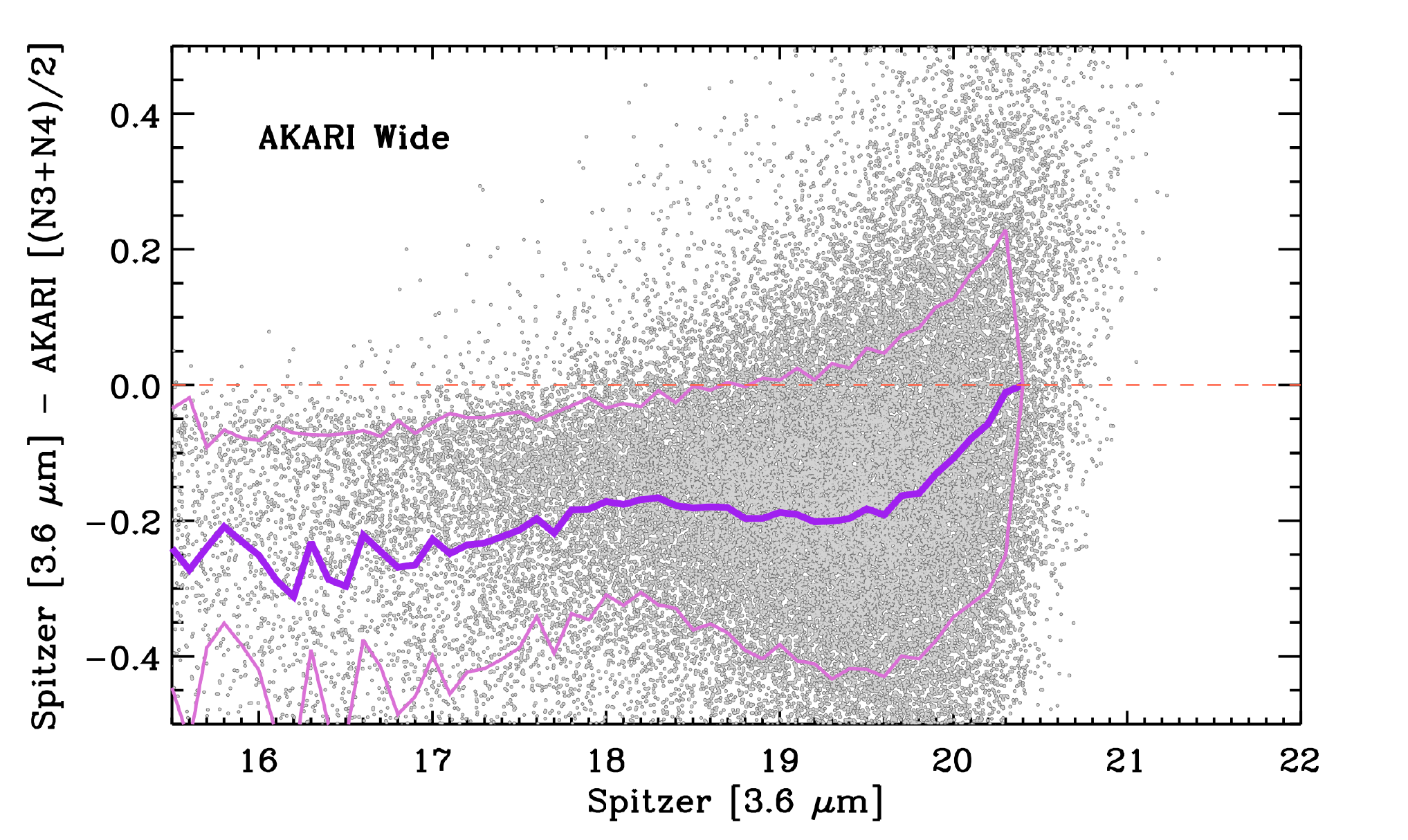}
\caption{Top: The photometric comparison of the {\it Spitzer} NEP to that
  of {\it WISE} \citep{Wright2010} for the NEP 3.6\,$\mu$m (left) and
  4.5\,$\mu$m bands. We use the AllWISE catalog \citep{Cutri2013}
  photometry measurements in the W1 (at 3.4\,$\mu$m) and W2 (at
  4.6\,$\mu$m) bands to compare with NEP catalog respectively. The NEP
photometry is consistent with {\it WISE} showing small scatter with
magnitude offsets associated with variations in the filter response
function shapes between {\it Spitzer} and {\it WISE}. Bottom:
Photometric comparison of the IRAC 3.6\,$\mu$m in the NEP to that of
the AKARI deep (left) \citep{Murata2013} and AKARI wide (right)
\citep{Kim2012}. The AKARI photometry is taken from the average of the
fluxes in the two AKARI infrared bands of N3 and N4 (at 3.2\,$\mu$m and
4.1\,$\mu$m respectively).}
\label{fig:Fig9}
\end{figure*}

\begin{figure}
\centering
\includegraphics[trim=2cm 0cm 0cm 0cm,scale=0.4]{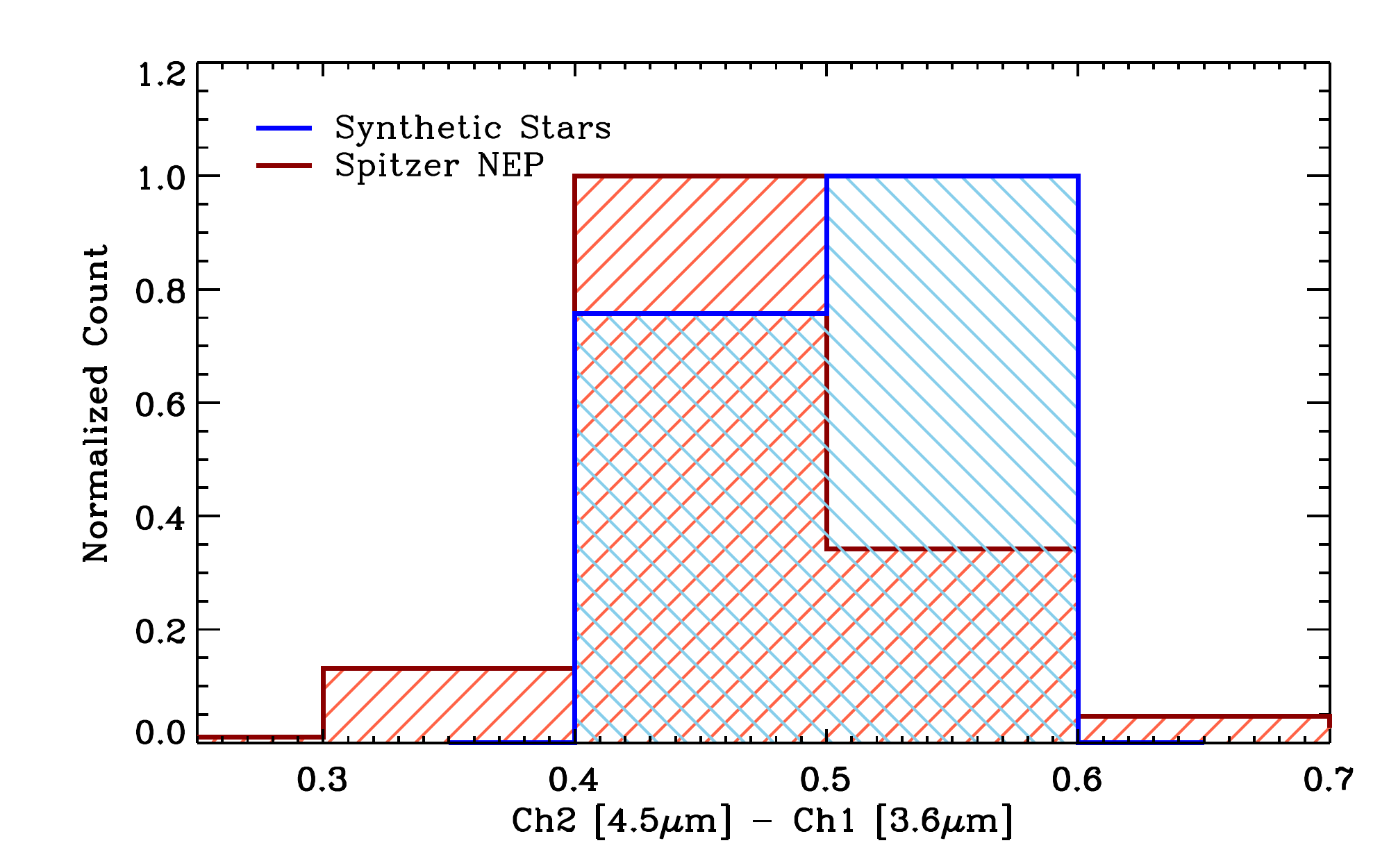}
\caption{The observed $\rm (4.5\,\mu m-3.6\,\mu m)$ color distributions of
  stars in the {\it Spitzer} NEP catalog (in red) compared to the
  color of template stellar models from {\sc BaSeL} library \citep{Lejeune1997, Lejeune1998,
  Westera2002}. The NEP stars are selected using the {\sc SExtractor}
stellarity parameter assuming {\sc class\_star}\,$>0.9$. The stellar
model colors are computed using the {\it Spitzer} filter response
functions. The observed IRAC color of stars is consistent with
predictions from the {\sc BaSeL} library, with mean colors of 0.54 and 0.51
respectively. The scatter in the observed IRAC color is associated
with the photometric uncertainties in the NEP catalog.}
\label{fig:Fig10}
\end{figure}

\begin{figure}
\centering
\includegraphics[trim=2cm 0cm 0cm 0cm,scale=0.4]{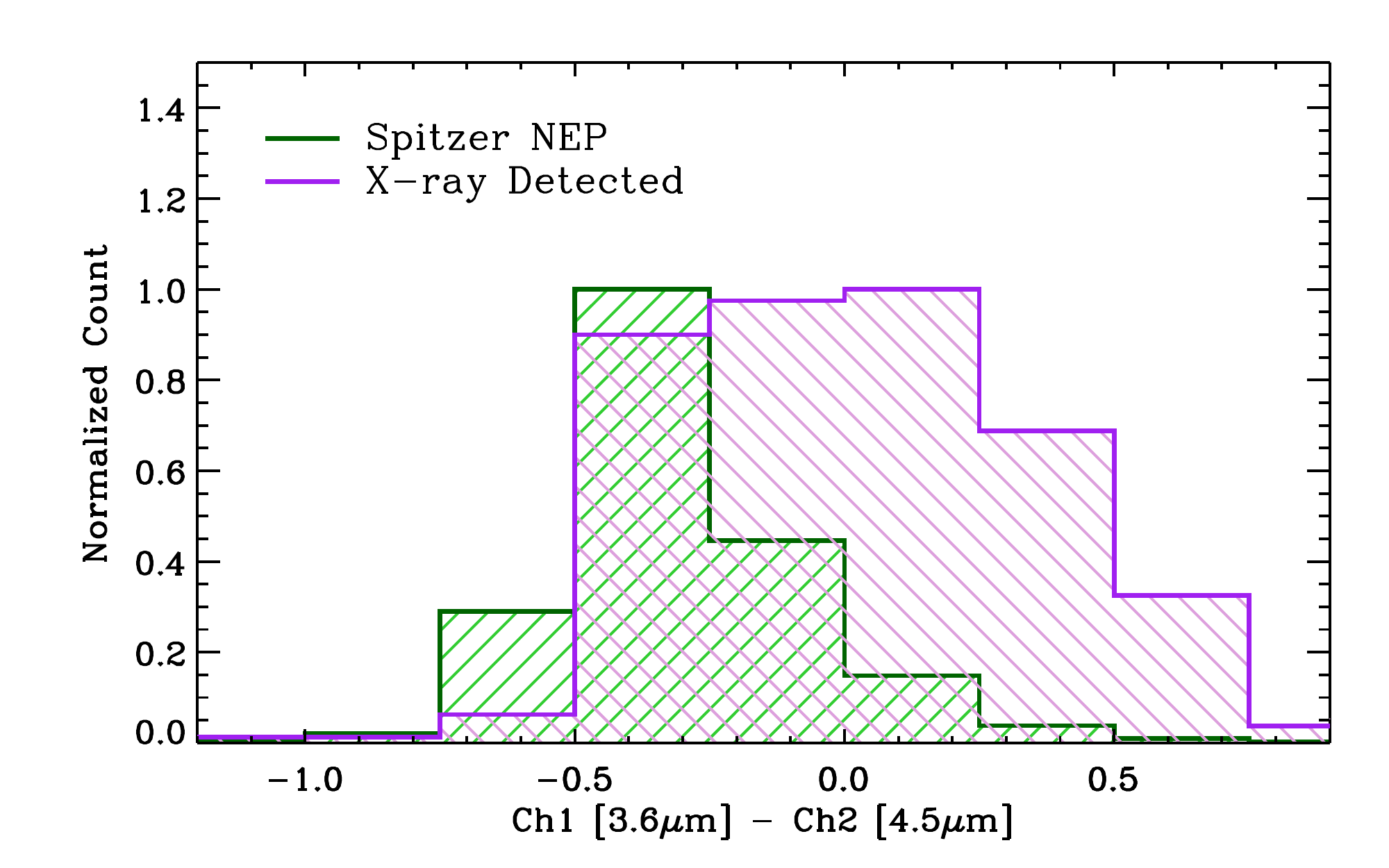}
\caption{The observed IRAC $\rm (3.6\,\mu m-4.5\,\mu m)$ color
  distribution of all {\it Spitzer} NEP sources (in green) compared to
  the X-ray detected objects \citep{Krumpe2015} (in magenta). The distribution
  of IRAC color of X-ray sources is consistent with the AGN colors
  reported in \citealp{Stern2005}.}
\label{fig:Fig11}
\end{figure}

\subsection{Aperture Corrections}

We perform aperture photometry over two circular apertures at fixed
$4^{\prime\prime}$ and $6^{\prime\prime}$ diameters in addition
to the {\sc SExtractor} measured {\sc mag\_auto} discussed above. The fixed
apertures often under-predict the flux of objects. 
We corrected the aperture photometry measurements in the 3.6\,$\mu$m
and 4.5\,$\mu$m bands for the effect of fixed apertures (at
$4^{\prime\prime}$ and $6^{\prime\prime}$) used in their
measurements. For this we used the curve of growth of five isolated
stars in the field using ever growing circular apertures (with
$4^{\prime\prime}-10^{\prime\prime}$ diameter) to measure the source
flux. The aperture correction is measured as the mean of the
correction from the curve of growth of the selected stars. Table 2
summarizes the aperture corrections (in AB magnitudes) for both observed
IRAC bands. These agree with the reported numbers by \citet{Ashby2009}
for the SDWFS and also corrections reported by the IRAC Instrument
Handbook.

\begin{table}
\begin{center}
\caption{Aperture Corrections$^\dagger$.}
\begin{tabular}{ccc}
\hline
\hline
Aperture Diameter & 3.6\,$\mu$m band & 4.5\,$\mu$m band  \\ 
\hline
$3^{\prime\prime}$ & -0.65 & -0.65 \\
$4^{\prime\prime}$ & -0.36 & -0.46 \\
$5^{\prime\prime}$ & -0.22 & -0.28 \\
$6^{\prime\prime}$ & -0.15 & -0.19 \\
\hline
\end{tabular}
\end{center}
\footnotesize
$^\dagger$: Measured from point-sources curve of growth.
\label{table:Table2}
\end{table}

\subsection{Astrometry}

We validate the astrometry of the {\it Spitzer} NEP catalog generated by {\sc
  SExtractor} against the publicly available USNO CCD Astrograph
Catalog (UCAC; \citealp{Zacharias2013}) fourth version\footnote{\url{http://www.usno.navy.mil/USNO/astrometry/optical-IR-prod/ucac}}. We use a 2$^{\prime\prime}$
radius to cross-match the NEP catalog with that of UCAC. We further
limit the astrometry analysis to sources with a minimum
signal-to-noise ratio of 20 in the NEP 3.6\,$\mu$m band. This
generates a catalog of 8729 sources over the full
NEP area. Figure \ref{fig:Fig7} shows the distribution of the
difference in right ascension and declination between the NEP catalog
and the UCAC. The distribution of the difference in declination is
consistent with being centered on zero with a deviation of
0$^{\prime\prime}$.04, whereas the right ascension difference shows a
mean offset of 0$^{\prime\prime}$.25. These distributions have
standard deviations of 0$^{\prime\prime}$.51 and 0$^{\prime\prime}$.24
in the right ascension and declination respectively, smaller than the
IRAC spatial resolution measured from the PSF FWHM ($\sim 1^{\prime\prime}.78$).  We note here
that the offset is also smaller than the pixel scale of our {\it Spitzer}/IRAC
mosaics at 1$^{\prime\prime}$.2 and that we did not apply this to the
final photometric catalog. The final photometric
catalog will be available through the IRSA
website\footnote{\url{http://irsa.ipac.caltech.edu/Missions/spitzer.html}}
and as a machine readable table with this publication. Table 3 in
the appendix summarizes the entries in the published catalog.

\section{Analysis}

\subsection{Number Counts}

The {\it Spitzer} NEP 3.6\,$\mu$m selected catalog has 380,858 sources
over an area of 7.04\,deg$^2$. Figure \ref{fig:Fig6} shows the NEP
area-normalized source number counts along with the source counts from
the {\it Spitzer} Deep Wide-Field Survey (SDWFS;
\citealp{Ashby2009}) in the 3.6\,$\mu$m and 4.5\,$\mu$m. The
variations in the number count distributions
are associated with the larger number of stars and shallower depths (see
Figure \ref{fig:Fig2}) at the bright and faint end respectively in the
NEP compared to that of SDWFS. Table 3 summarizes the source counts
of the photometric catalog in designated magnitude bins along with the
associated Poisson uncertainties.

\begin{table}
\begin{center}
\caption{{\it Spitzer}/IRAC 3.6\,$\mu$m number counts in the NEP catalog.}
\begin{tabular}{*{3}{c}}
\hline
\hline
{\it Spitzer}/IRAC 3.6\,$\rm \mu m^{\dagger}$ & N ($\text{deg}^{-2}\text{mag}^{-1}$) & Poisson Uncertainty \\
\hline

       14.25&        121&            6 \\
       14.75&        170&            7 \\
       15.25&        259&            9 \\
       15.75&        404&           11 \\
       16.25&        586&           13 \\
       16.75&        845&           15 \\
       17.25&       1270&           19 \\
       17.75&       1927&           23 \\
       18.25&       3222&           30 \\
       18.75&       5522&           40 \\
       19.25&       8605&           49 \\
       19.75&      11811&           58 \\
       20.25&      14422&           64 \\
       20.75&      15205&           66 \\

\hline
\end{tabular}
\end{center}
\footnotesize $^{\dagger}$: Bin center magnitude.
\label{table:Table2}
\end{table}

\subsection{Completeness Estimates}

To get an estimate of the NEP survey catalog depth we performed a
point-source completeness estimate on the combined three epoch full depth in
the 3.6\,$\mu$m band. For this we generated a large
number of point sources of varying brightness (with 1000 sources in
each 0.05 bins of magnitude), given the IRAC PSF
FWHM (at $1^{\prime\prime}.78$). We then randomly distributed these
sources across the 3.6\,$\mu$m mosaic while avoiding the mosaic edges
and existing objects in the field using the segmentation map generated
by {\sc SExtractor} for the main photometric catalog. We used
the original {\sc SExtractor} detection criteria to identify sources
in the new co-added map of real and simulated objects. We
cross-matched the generated catalog with the known simulated positions
of the injected sources which yields a recovery fraction of the point
sources. Figure \ref{fig:Fig8} shows the recovered fraction of the
simulated point sources as a function of the source brightness. The
catalog has a point source completeness estimate 80\% for objects with
19.71 (AB mag) in the 3.6\,$\mu$m and drops rapidly for faint
sources.

\subsection{Photometric Validation Check}

To check the robustness of the {\it Spitzer} measured photometry, we
compared our IRAC measured photometry with that of the Wide-Field
Infrared Survey Explorer ({\it WISE}; \citealp{Wright2010}). For this, we used the
AllWISE catalog \citep{Cutri2013} to extract the {\it WISE} photometry in
the W1 and W2 bands (at 3.4\,$\mu$m and 4.6\,$\mu$m respectively). We
used the conversion factors provided in Table 1 of \citet{Jarrett2011} to convert the {\it WISE} measured
Vega magnitudes in the W1 and W2 bands to AB magnitudes for comparison. Figure
\ref{fig:Fig9} shows the comparison of {\it
  Spitzer} measured magnitudes in the 3.6\,$\mu$m and 4.5\,$\mu$m
bands to that of {\it WISE} W1 and W2 observations in the NEP. The
{\it Spitzer} measured photometry in consistent with {\it WISE} with a
small scatter at bright magnitudes ($\rm \sigma(m) =0.03$ for 3.6\,$\mu$m
and $\rm \sigma(m) =0.01$ for 4.5\,$\mu$m; computed at $\rm m_{AB}=15$)
increasing to $\rm \sigma(m) =0.23$ for 3.6\,$\mu$m and $\rm \sigma(m)
=0.32$ for 4.5\,$\mu$m for fainter objects (measured at $\rm m_{AB}=20$). The {\it Spitzer} photometry is offset from {\it WISE} by $\rm \Delta m =0.13$ and $\rm
\Delta m =0.07$ in the  3.6\,$\mu$m and 4.5\,$\mu$m
respectively (measured from the bright end). This is associated with
differences in the filter response functions and effective wavelengths
probed creating a zero-point offset. We do not apply these offsets to
the our {\it Spitzer} NEP photometric catalog.

We further compared our IRAC photometry in the {\it Spitzer} NEP with infrared observations
by the AKARI in the north ecliptic pole. For this we used the catalog
of sources in the AKARI deep and wide surveys of the north ecliptic
pole \citep{Murata2013, Kim2012}, covering an area of 0.5\,deg$^2$ and
5.4\,deg$^2$ respectively. Figure \ref{fig:Fig9} shows the comparison
of the 3.6\,$\mu$m photometry to that of AKARI. For the comparison, we
take the average of the fluxes in the AKARI N3 and N4 filters (at 3.2\,$\mu$m
and 4.1\,$\mu$m respectively; \citealp{Murata2013}). We see from the figure that, on
average, the measured IRAC 3.6\,$\mu$m photometry agrees well with
that of AKARI, specially at the bright end and for the AKARI deep
observations. The main source of the deviation is associated with the
differences in the {\it Spitzer} filter response function and the
average filter functions associated with the combined AKARI N3 and N4 observations. 

\subsection{IRAC Color}

We further check the validity of the {\it Spitzer} measured
photometry by comparing the color distribution of stars in the NEP survey
to the expected infrared color derived from stellar model templates. For this
we used the {\sc BaSeL} stellar library \citep{Lejeune1997, Lejeune1998,
  Westera2002} and computed the expected color of model templates by
integrating over the stellar SEDs given the {\it Spitzer}/IRAC filter
response functions\footnote{\url{http://irsa.ipac.caltech.edu/data/SPITZER/docs/irac/calibrationfiles/spectralresponse/}} at 3.6\,$\mu$m and 4.5\,$\mu$m. Figure
\ref{fig:Fig10} shows the expected $\rm Ch2\,[4.5\,\mu m]-Ch1\,[3.6]\,\mu
m$ color of stellar models compared to measured colors of stars in the
NEP catalog. We identified stellar sources in the NEP catalog using
the stellarity parameter, {\sc class\_star}, from {\sc SExtractor}
requiring {\sc class\_star}\,$>0.99$. This gives 2107 stellar objects
for the NEP catalog. The IRAC color distribution of stars in the
NEP catalog is consistent with stellar model predictions with mean
$\rm (4.5\,\mu m-3.6\,\mu m)$ values of 0.47 and 0.50 for the observed
color of stars from NEP catalog and measured color from template
stellar models respectively. The scatter in the NEP IRAC colors is
associated with the photometric uncertainties in individual flux measurements.   

\subsection{X-ray Sources}

The {\it Spitzer}/IRAC colors have been used to identify
populations of active galaxies \citep{Stern2005, Donley2012} where the
continuum flux is expected to be dominated by a power law for AGN
dominated sources. These selections use the IRAC observations in four
filters to isolate the AGN selection area on the color-color
diagram. Due to lack of longer wavelength {\it Spitzer} observations
(at 5.8\,$\mu$m and 8.0\,$\mu$m), we instead look at the $\rm
(3.6\,\mu m-4.5\,\mu m)$ color distribution of sources in the NEP
catalog. Figure \ref{fig:Fig11} shows the IRAC color distribution of
all the NEP sources compared to the X-ray detected sources in the
field. We used the Chandra AKARI NEP X-ray point source catalog
\citep{Krumpe2015} and used a 2$^{\prime\prime}$ matching radius to
extract the IRAC colors. This gives 354 matched sources with the X-ray
catalog. We see that the X-ray sources have a redder $\rm
(3.6\,\mu m-4.5\,\mu m)$ color distribution, on average, compared to the full
sample as expected for AGN samples \citep{Stern2005}.

\section{Summary}

Here we presented a {\it Spitzer}/IRAC combined mosaics and a
3.6\,$\mu$m and 4.5\,$\mu$m detected catalog of extra-galactic sources
in the North Ecliptic Pole (NEP; centered
at $\rm R.A.=18^h00^m00^s$, $\rm Decl.=66^d33^m38^s.552$). The catalog
contains IRAC photometry for 380,858 sources in the 3.6\,$\mu$m and
4.5\,$\mu$m bands. Here are the main findings:

\begin{itemize}

\item The combined three epoch full-depth mosaic covers an area of
  7.04\,deg$^2$ reaching 1$\sigma$ depths of
  1.29\,$\mu$Jy and 0.79\,$\mu$Jy, over an aperture of
  4$^{\prime\prime}$ in diameter, in the 3.6\,$\mu$m and
    4.5\,$\mu$m bands respectively.

\item The mosaics have uniform exposure across the field with more
  than 89\% of the field covered by an exposure time of at least
  100\,sec.

\item The photometric catalog has astrometry offset and uncertainties
  consistent with the {\it Spitzer} PSF FWHM and pixel size. 

\item The {\it Spitzer}/IRAC measured photometry in the NEP is consistent
  with earlier measurements by {\it WISE} and AKARI at similar wavelengths with
  current IRAC observations being deeper than both.

\item The colors of stellar objects in the NEP agrees with colors of
  stellar template models and also agrees with expected colors of
  X-ray detected sources, further validating the measured photometry.

\item The generated IRAC point-source catalogs will be available
  through the NASA/IPAC Infrared Science Archive.

\end{itemize}

\section*{Acknowledgement} 

We wish to thank the referee for reading the original manuscript
  and providing useful suggestions. Financial support for this work was provided by NSF
through AST-1313319 for H.N. and A.C. UCI group also acknowledges
support from HST-GO-14083.002-A, HST-GO-13718.002-A and NASA
NNX16AF38G grants. M.I. acknowledge the support from the NRFK grant,
No. 2017R1A3A3001362. This work was supported by NASA APRA research
grants NNX07AI54G, NNG05WC18G, NNX07AG43G, NNX07AJ24G, and
NNX10AE12G. This work is based on observations made with the Spitzer
Space Telescope, which is operated by the Jet Propulsion Laboratory,
California Institute of Technology under a contract with NASA. Support
for this work was provided by NASA through an award issued by JPL/Caltech.

\bibliographystyle{apj}
\bibliography{references}

\newpage

\section*{Appendix: Photometric Catalog Entries}
The entries in the {\it Spitzer}/IRAC photometric catalog for the
North Ecliptic Pole. 

\vspace{1cm}

\noindent \# 1  ID \\
\# 2  R.A. \\
\# 3  Decl. \\
\# 4  IRAC 3.6\,$\mu$m Auto Magnitude (AB) \\
\# 5  IRAC 3.6\,$\mu$m Auto Magnitude Error (AB) \\
\# 6  IRAC 4.5\,$\mu$m Auto Magnitude (AB) \\
\# 7  IRAC 4.5\,$\mu$m Auto Magnitude Error (AB) \\
\# 8  IRAC 3.6\,$\mu$m Aperture1 Magnitude (AB) \\
\# 9  IRAC 3.6\,$\mu$m Aperture1 Magnitude Error (AB) \\
\# 10  IRAC 4.5\,$\mu$m Aperture1 Magnitude (AB) \\
\# 11  IRAC 4.5\,$\mu$m Aperture1 Magnitude Error (AB) \\
\# 12  IRAC 3.6\,$\mu$m Aperture2 Magnitude (AB) \\
\# 13  IRAC 3.6\,$\mu$m Aperture2 Magnitude Error (AB) \\
\# 14  IRAC 4.5\,$\mu$m Aperture2 Magnitude (AB) \\
\# 15  IRAC 4.5\,$\mu$m Aperture2 Magnitude Error (AB) \\
\# 16  CLASS\_STAR\\

\noindent Notes: \\
Col. (1): Sequential number. \\
Col. (2) \& (3): Target coordinates (in degrees). \\
Col. (4) - (7): 3.6\,$\mu$m and 4.5\,$\mu$m photometry over Kron radius (AB mag)\\
Col. (8) - (11): 3.6\,$\mu$m and 4.5\,$\mu$m aperture photometry with
a $4^{\prime\prime}$ diameter aperture (AB mag)\\
Col. (12) - (15): 3.6\,$\mu$m and 4.5\,$\mu$m aperture photometry with
a $6^{\prime\prime}$ diameter aperture (AB mag)\\
Col. (16): {\sc SExtractor} stellarity parameter. \\

\end{document}